\documentclass[a4paper,11pt]{article}

\usepackage{jcappub}
\usepackage[utf8]{inputenc}
\usepackage{graphicx,color,amssymb,amsmath,mathtools,cases,bm,float,tocbibind,upgreek,longtable,geometry,blindtext,hyperref}
\usepackage[format=hang,font=small,textfont=it]{caption}
\allowdisplaybreaks[2]

\usepackage{pifont}  
\usepackage{subcaption}

\title{Numerical Study of Wheeler-Dewitt Equation beyond Slow-roll approximation}
\author[a,b,1]{Jie Jiang,\note{Corresponding author}}
\author[a,b]{Deog Ki Hong,}
\author[c,d,e,f]{Dong-han Yeom}

\affiliation[a]{Extreme Physics Institute, Pusan National University, \\Busan 46241, Republic of Korea}
\affiliation[b]{Department of Physics, Pusan National University, \\Busan 46241, Republic of Korea}
\affiliation[c]{Department of Physics Education, Pusan National University, \\Busan 46241, Republic of Korea}
\affiliation[d]{Research Center for Dielectric and Advanced Matter Physics, Pusan National University, \\Busan 46241, Republic of Korea}
\affiliation[e]{Leung Center for Cosmology and Particle Astrophysics,
National Taiwan University, \\Taipei 10617, Taiwan}
\affiliation[f]{Department of Physics and Astronomy, University of Waterloo, Waterloo, ON N2L 3G1, Canada}
\date{}
\emailAdd{jiejiang@pusan.ac.kr}
\emailAdd{dkhong@pusan.ac.kr}
\emailAdd{innocent.yeom@gmail.com}

\abstract{
The Wheeler-DeWitt (WDW) equation is analyzed using two boundary proposals: the Hartle-Hawking no-boundary condition and tunneling condition. By compactifying the scale factor $a$ into $ x = a/(1+a) $, we reformulate the WDW equation to find stable numerical solutions with clearer boundary conditions. The no-boundary wave function peaks at the horizon scale, indicating quantum nucleation of classical spacetime, while the tunneling solution shows exponential decay, reflecting vacuum decay from a classically forbidden state. These dynamics are explored under slow-roll and non-slow-roll regimes of a periodic potential, separately, with non-slow-roll scenarios amplifying quantum effects that delay the classical behavior. The results emphasize the role of boundary conditions in quantum cosmology, offering insights into the universe's origin and the interplay between quantum gravity and observable cosmology.
}

\begin{document}
\hfill {\parbox[b]{1in}{ \hbox{\tt PNUTP-25/A03}  }

\maketitle


\section{Introduction}

The Wheeler-DeWitt (WDW) equation stands as a cornerstone of quantum cosmology, offering a framework to describe the quantum state of the entire universe \cite{DeWitt:1967yk}. Arising from the canonical quantization of Einstein’s general relativity, it governs the wave function $\psi(a, \phi)$ of the universe, where $a$ is the scale factor and $\phi$ represents matter fields. Unlike conventional quantum mechanics, the WDW equation operates in superspace—the infinite-dimensional space of all possible 3-geometries and field configurations—making it a unique challenge in theoretical physics. Its solutions, interpreted through boundary conditions, provide insights into the universe’s primordial quantum origins and the emergence of classical spacetime.

Two prominent boundary conditions dominate this discourse: the Hartle-Hawking no-boundary proposal \cite{Hartle:1983ai} and Vilenkin’s tunneling proposal \cite{Vilenkin:1984wp}. The no-boundary condition posits a smooth transition from an Euclidean quantum state to a Lorentzian universe, avoiding initial singularities. Conversely, the tunneling proposal envisions the universe nucleating via quantum tunneling from `nothing' (a classically forbidden region). These competing visions reflect deeper philosophical divides about the nature of time, causality, and the role of observers in quantum cosmology.

One can contrast two wave functions in many aspects. The no-boundary proposal is constructed as the Euclidean path integral, which is approximated by complexified instantons \cite{Hartle:2008ng}. Although only a small amount of inflation is preferred in this wave function \cite{Hwang:2013nja}, there might be a way to enhance a large number of $e$-foldings based on several mechanisms \cite{Hwang:2012bd,Maldacena:2024uhs}. On the other hand, the result of the tunneling proposal might be explained from the Lorentzian path integral \cite{Feldbrugge:2017kzv}. The tunneling proposal typically prefers initial conditions for large $e$-foldings, which might be helpful to find consistent inflation models in the context of swampland conjectures \cite{Brahma:2020cpy}.

What is the intuitive reason that two wave functions provide different results \cite{Vilenkin:1987kf}? With the ultra-slow-roll approximation, the no-boundary wave function assumes only the increasing modes in the quantum domain; this results in the superposition of outgoing and ingoing modes in the classical domain. On the other hand, the tunneling wave function assumes only the outgoing modes in the classical domain; this requires a superposition of increasing and decreasing modes in the quantum domain. Since the quantum domain mainly determines the probability, the no-boundary wave function is determined by the increasing modes, while the decreasing modes dominate the tunneling wave function. In terms of the WKB approximation, this manifests in the \textit{sign difference} in front of the action; hence, it is not very surprising that the no-boundary wave function is evaluated by the Euclidean path integral, while the tunneling proposal is related to the Lorentzian path integral, although there might be several unclear issues.

However, historically, almost all discussions were restricted to the ultra-slow-roll approximation. If we turn on the dynamics in the field direction, the boundary condition problem is not simple any more. How do we provide the boundary conditions along the field direction? Also, what happens if we cannot ignore the field derivatives in the WDW equation? Up to now, the best strategy has been to look for complexified instantons for non-slow-roll cases \cite{Hwang:2011mp,Yeom:2021twr}; however, one can ask whether the instanton approximation is still good for non-slow-roll cases. Also, it is a challenging question: how can we provide the increasing modes only for the small-scale factor regime (no-boundary) or the outgoing modes only for the large-scale factor regime (tunneling) in the numerical framework? One may get some help from the WKB approximation \cite{Kang:2022tkb}, but we need more careful treatment to compare the two proposals.

In this work, we revisit these proposals through a novel lens: a coordinate transformation that reparametrizes the scale factor $a$ into a compactified variable $x = a/(1+a)$. This transformation maps the semi-infinite domain $a \in [0, \infty)$ to a finite one, $x \in [0, 1)$, simplifying the imposition of boundary conditions and enhancing the stability in numerical analysis. By solving the transformed WDW equation under both proposals, we analyze the interplay between quantum tunneling, semiclassical transitions, and potential parameters in a periodic potential landscape.

Our results reveal distinct signatures of each proposal: the no-boundary wave function peaks at the horizon scale, signaling quantum nucleation, while the tunneling solution exhibits exponential decay characteristic of metastable vacuum decay. These findings could deepen our understanding of quantum gravity’s conceptual framework and its observational implications.

\section{Wheeler-DeWitt Equation}

The Wheeler-DeWitt equation is a fundamental result in the Hamiltonian formulation of general relativity, providing a framework for quantum cosmology \cite{Kang:2022tkb}. It arises from the canonical quantization of the Einstein-Hilbert action to describe the dynamics of a homogeneous and isotropic universe containing a scalar field $\phi$. The Einstein-Hilbert action, including the scalar field, is given by, taking the Newton's constant $G_N=1$:

\begin{equation}
	S = \int d^4x \sqrt{-g} \left[ \frac{\mathcal{R}}{16\pi} - \frac{1}{2} (\partial \phi)^2 - V(\phi) \right] ~ ,
\end{equation}

\noindent where $\mathcal{R}$ is the Ricci scalar, $g$ is the determinant of the metric tensor, $(\partial \phi)^2$ represents the kinetic term of the scalar field, and $V(\phi)$ is the potential energy associated with the field.

To simplify the analysis, we consider a closed Friedmann-Robertson-Walker (FRW) metric, which describes a spatially homogeneous and isotropic universe. The metric is expressed in terms of a lapse function $N(t)$ and the scale factor $a(t)$ as follows: 

\begin{equation}
	ds^2 = -N^2(t) dt^2 + a^2(t) d\Omega_3^2 ~ ,
\end{equation}

\noindent where $d\Omega_3^2$ is the metric on a three-dimensional sphere. Under this ansatz, the action reduces to a one-dimensional minisuperspace model, which simplifies the dynamics to a single degree of freedom governed by the scale factor $a(t)$ and the scalar field $\phi(t)$. The reduced action takes the form:

\begin{equation}
	S = 2\pi^2 \int dt N \left[ \frac{3}{8\pi} \left( -\frac{a \dot{a}^2}{N^2} + a \right) + \frac{1}{2} a^3 \frac{\dot{\phi}^2}{N^2} - a^3 V(\phi) \right] ~ .
    \label{action}
\end{equation}
Here, the first term represents the gravitational contribution, while the second and third terms describe the kinetic and potential energy of the scalar field, respectively.
From this action, Eq.~(\ref{action}), we derive the canonical momenta conjugates to the scale factor $a$ and the scalar field $\phi$, respectively:

\begin{equation}
\begin{aligned}
    p_a &= -\frac{3\pi a \dot{a}}{2N}, \\
    p_{\phi} &= \frac{2\pi^2 a^3 \dot{\phi}}{N}.
\end{aligned}
\end{equation}

These momenta are essential for constructing the Hamiltonian of the system. The Hamiltonian constraint, which arises due to the time-reparameterization invariance of general relativity, is given as 

\begin{equation}
    H = N \left[ -\frac{p_a^2}{3\pi a} + \frac{p_{\phi}^2}{4\pi^2 a^3} - \frac{3\pi a}{4} + 2\pi^2 a^3 V(\phi) \right] = 0.
\end{equation}
This constraint reflects the fact that the total energy of the universe, including both gravitational and matter contributions, must vanish.

To transit to the quantum regime, we replace the classical momenta with their corresponding quantum operators. This is achieved by the canonical quantization procedure, where 

\begin{equation}
\begin{aligned}
    p_a &\rightarrow -i \frac{\partial}{\partial a}, \\
    p_{\phi} &\rightarrow -i \frac{\partial}{\partial \phi}.
\end{aligned}
\end{equation}
Substituting these operators into the Hamiltonian constraint yields the Wheeler-DeWitt equation, a hyperbolic partial differential equation that governs the quantum dynamics of the universe in minisuperspace:

\begin{equation}\label{WDE}
    \left[ \frac{1}{a} \frac{\partial^2}{\partial a^2} - \frac{1}{a^2} \frac{\partial}{\partial a} - \frac{3}{4\pi a^3} \frac{\partial^2}{\partial \phi^2} - \frac{9\pi^2 a}{4} + 6\pi^3 a^3 V(\phi) \right] \psi(a, \phi) = 0.
\end{equation}

\noindent Here, $\psi(a, \phi)$ is the wave function of the universe, which depends on the scale factor $a$ and the scalar field $\phi$. The WDW equation describes the quantum behavior of the universe in a simplified minisuperspace model, where the Laplace-Beltrami operator is chosen for mathematical simplicity \cite{Kiefer:2007ria}. This equation plays a central role in quantum cosmology, offering insights into the quantum nature of spacetime and the early universe.

\section{Boundary condition}

\subsection{No-boundary proposals}

Hartle and Hawking’s no-boundary proposal defines the wave function as a Euclidean path integral over compact geometries without boundaries \cite{Hartle:1983ai}:

\begin{equation}
	\psi(\tilde{g}, \tilde{\phi}) = \int \mathcal{D} g \mathcal{D} \phi \, e^{- S_{\mathrm{E}} (g, \phi)} ~ ,
\end{equation}
where this integration sums over all regular and compact geometries that has the boundary values $\tilde{g}$ and $\tilde{\phi}$.

In the WKB approximation, the wave function transitions between exponential and oscillatory behaviors at the horizon scale $a_H = 1/\mathcal{H}(\phi)$, where $\mathcal{H}(\phi) = \sqrt{8\pi V(\phi)/3}$ \cite{Kiefer:2007ria}:
In the auantum regime ($a < a_H$) it becomes

\begin{equation}
    \psi(a, \phi) \propto \frac{1}{(1 - \mathcal{H}^2 a^2)^{1/4}} \exp\left[\frac{\pi}{2\mathcal{H}^2}\left(1 - (1 - \mathcal{H}^2 a^2)^{3/2}\right)\right]\,,
\end{equation}
while in the 
classical regime ($a > a_H$) one finds
\begin{equation}
    \psi(a, \phi) \propto \frac{1}{(\mathcal{H}^2 a^2 - 1)^{1/4}} \exp\left(\frac{\pi}{2\mathcal{H}^2}\right) \exp\left[\frac{i\pi}{2\mathcal{H}^2}(\mathcal{H}^2 a^2 - 1)^{3/2}\right].
\end{equation}
The oscillatory phase in the classical refime correlates with classical trajectories, embodying the emergence of Lorentzian spacetime from a Euclidean quantum state.

\subsection{Tunneling proposals}

Vilenkin’s tunneling proposal interprets the wave function as a outgoing flux emanating from $a = 0$, which represents quantum nucleation from a classically forbidden region ($a < a_H$) \cite{Kiefer:2007ria}. The probability current,

\begin{equation}
    j = \frac{i}{2} \left( \psi^* \nabla \psi - \psi \nabla \psi^* \right),
\end{equation}

\noindent dictates boundary conditions, requiring purely outgoing waves at $a \to \infty$. The wave function exhibits two characteristic features: 
For $a < a_H$, it decays exponentially 
\begin{equation}
    \psi(a, \phi) \propto \frac{1}{(1 - \mathcal{H}^2 a^2)^{1/4}} \exp\left[-\frac{\pi}{2\mathcal{H}^2}\left(1 - (1 - \mathcal{H}^2 a^2)^{3/2}\right)\right]\,,
\end{equation}
while for $a > a_H$ it corresponds to 
outgoing waves, 
\begin{equation}
    \psi(a, \phi) \propto \frac{1}{(\mathcal{H}^2 a^2 - 1)^{1/4}} \exp\left(-\frac{\pi}{2\mathcal{H}^2}\right) \exp\left[\frac{i\pi}{2\mathcal{H}^2}(\mathcal{H}^2 a^2 - 1)^{3/2}\right].
\end{equation}
This contrasts with the no-boundary condition’s time-symmetric structure, favoring a universe that tunnels into the current one rather than "smoothly" emerging.

\section{Transformation of Wheeler-DeWitt equation}

To circumvent the numerical challenges posed by the infinite domain of $a$, we introduce a compactified variable:

\begin{equation}
    x = \frac{a}{1 + a}, 
\end{equation}

\noindent mapping $a \in [0, \infty)$ to $x \in [0, 1)$. In terms of the new variable, Eq.~\eqref{WDE} becomes

\begin{equation}\label{WDEx}
    \psi_{,xx} - \frac{3}{4 \pi (x - 1)^2 x^2} \psi_{,\phi\phi} + \frac{(1 + 2x)}{(x - 1)x} \psi_{,x} - \frac{3\pi^2 x^2 \left( 3 (x - 1)^2 - 8\pi x^2 V(\phi) \right)}{4 (x - 1)^8} \psi = 0.
\end{equation}

This reformulation offers two advantages: Firstly,  it facilitates discretization and the imposition of boundary conditions. Secondly, it removes the singularities at infinity, $a \to \infty$.

We impose boundary conditions at the limits $ x = 0 $ and $ x = 1 $. At these boundaries, we assume that the wave function exhibits no explicit field dependence, allowing us to utilize the WKB approximation in terms of the variable $ a $. The boundary conditions must be therefore reformulated accordingly in the $ x $ variable.

\subsection{No-boundary Proposal}

For quantum regime ($ x \sim 0 $), the wave function takes the following form

\begin{equation}
    \psi(x, \phi) \propto \frac{1}{\left[ 1 - \frac{\mathcal{H}^2 x^2}{(1 - x)^2} \right]^{1/4}} \exp \left[ \frac{\pi}{2\mathcal{H}^2} \left( 1 - \left( 1 - \frac{\mathcal{H}^2 x^2}{(1 - x)^2} \right)^{3/2} \right) \right].
\end{equation}
Similarly, for classical regime ($ x \sim 1 $), we obtain

\begin{equation}
    \psi(x, \phi) \propto \frac{1}{\left[ \frac{\mathcal{H}^2 x^2}{(1 - x)^2} - 1 \right]^{1/4}} \exp \left( \frac{\pi}{2\mathcal{H}^2} \right) \exp \left[ \frac{i\pi}{2\mathcal{H}^2} \left( \frac{\mathcal{H}^2 x^2}{(1 - x)^2} - 1 \right)^{3/2} \right].
\end{equation}

This formulation aligns with the Hartle-Hawking no-boundary proposal, where the universe emerges smoothly from a quantum state without an initial singularity.

\subsection{Tunneling Proposal}

For exponential decay ($ x \sim 0 $), the wave function follows
\begin{equation}
    \psi(x, \phi) \propto \frac{1}{\left[ 1 - \frac{\mathcal{H}^2 x^2}{(1 - x)^2} \right]^{1/4}} \exp \left[ - \frac{\pi}{2\mathcal{H}^2} \left( 1 - \left( 1 - \frac{\mathcal{H}^2 x^2}{(1 - x)^2} \right)^{3/2} \right) \right].
\end{equation}
For outgoing waves ($ x \sim 1 $), we obtain
\begin{equation}
    \psi(x, \phi) \propto \frac{1}{\left[ \frac{\mathcal{H}^2 x^2}{(1 - x)^2} - 1 \right]^{1/4}} \exp \left( - \frac{\pi}{2\mathcal{H}^2} \right) \exp \left[ \frac{i\pi}{2\mathcal{H}^2} \left( \frac{\mathcal{H}^2 x^2}{(1 - x)^2} - 1 \right)^{3/2} \right].
\end{equation}
This corresponds to the Vilenkin tunneling proposal, wherein the universe emerges via a quantum tunneling process from a classically forbidden state.

\section{Result}

\subsection{ A model with a periodic Potential }

We adopt a periodic potential, given as

\begin{equation}
	V ( \phi ) = V _ { 0 } + m \left( 1 + \cos \frac { \pi } { \Delta } \phi \right) ~ ,
\end{equation}

\noindent motivated by axion-like fields in inflationary cosmology. The parameters $V_0$ (vacuum energy), $m$ (modulation amplitude), and $\Delta$ (barrier width) control the landscape’s structure. This choice allows exploration of both slow-roll ($V_0 > m$) and non-slow-roll ($V_0 < m$) regimes.

\subsection{Slow-roll Regime ($V_0 = 0.2, m = 0.1, \Delta = 3$)}

\subsubsection{No-boundary Condition}

To address the divergence of the wave function $ \psi(x) $ at $ x = 0 $, we impose a regularization procedure by setting the no-boundary condition at $ x = 0.01 $, slightly off from the singularity. This approach is analogous to the Hartle-Hawking "no-boundary" proposal in quantum cosmology, where the wave function of the universe is defined by a path integral over compact geometries without singular initial conditions \cite{Hartle:2008ng}. By evolving $ \psi(x) $ forward from $ x = 0.01 $, we ensure physically interpretable solutions while mitigating numerical artifacts near the origin.  

The resultant wave function, shown in Fig.~\ref{hh_forward_sr}, the black curve is the location of the maximum value of the wave function for a given $\phi$, the blue curve denotes the expected analytical values at the horizon.  In Fig.~\ref{hh_forward_2d_sr}, we present the two curve, a slice in two-dimensions. The figures exhibit two distinct regimes:

1. Quantum Regime ($ a < \mathcal{H} $): Near the horizon $ a = \mathcal{H} $, $ \psi(x) $ reaches a local maximum, indicating a heightened probability amplitude for the system to reside in this region. This peak aligns with the critical point of the potential $ V(\phi) $, where quantum effects dominate.  

2. Classical Regime ($ a > \mathcal{H} $): Beyond the horizon, $ \psi(x) $ decays and transits into oscillatory behavior, characteristic of wave propagation in classically allowed regions. These oscillations reflect the system’s tendency to explore phase space trajectories consistent with classical dynamics, as predicted by the correspondence principle.  

The sharp decline at $ a = \mathcal{H} $ suggests a boundary between quantum and classical descriptions, emphasizing the role of the potential’s structure that causes this transition. Further analysis of the oscillation frequency and decay rate could elucidate the system’s semiclassical limit and its sensitivity to $ V_0 $ and $ \Delta $.

\begin{figure}
    \centering
    \begin{subfigure}{0.48\textwidth}
        \includegraphics[width=\textwidth]{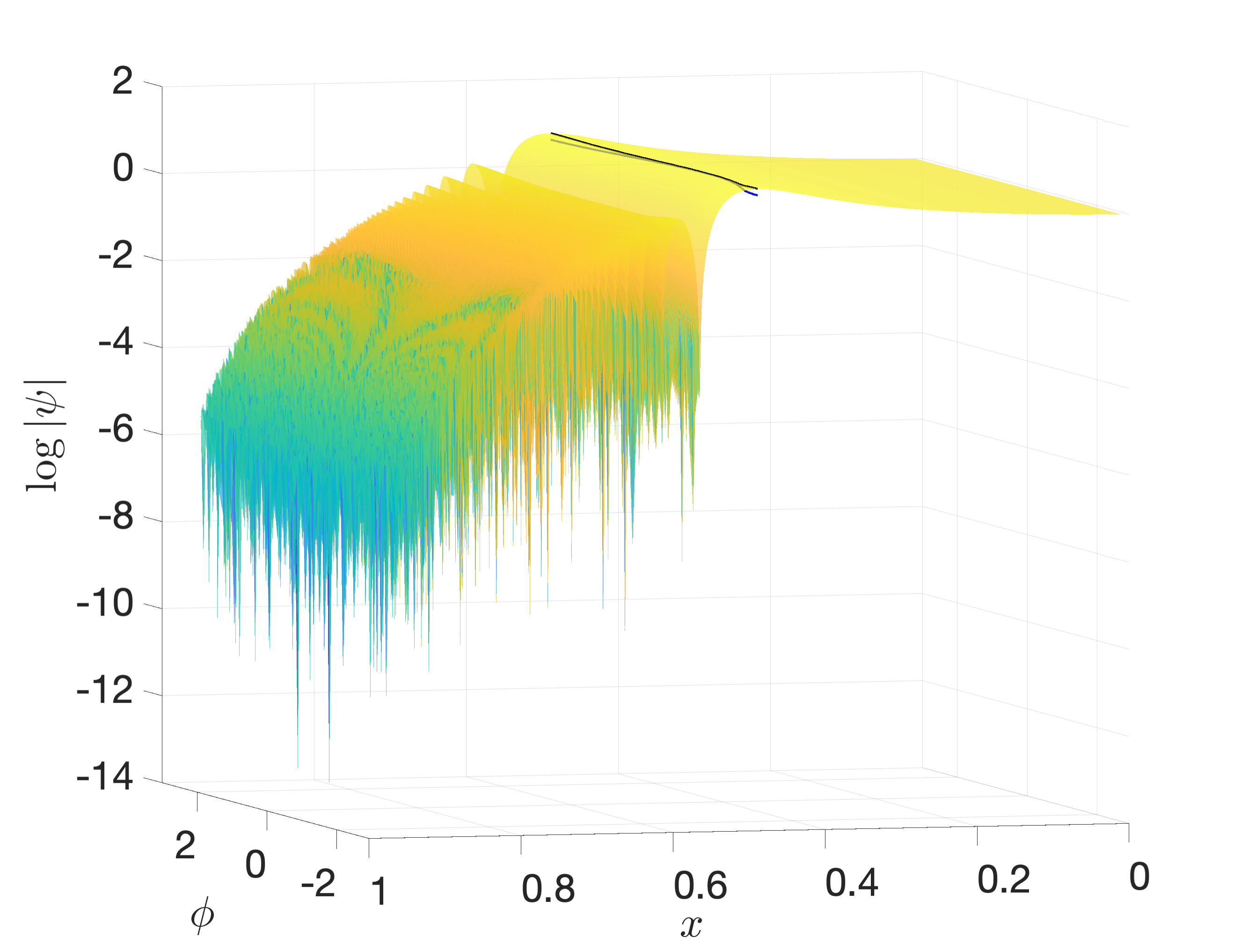}
        \caption{3 dimensional figure.}
        \label{hh_forward_sr}
    \end{subfigure}
    \hfill
    \begin{subfigure}{0.48\textwidth}
        \includegraphics[width=\textwidth]{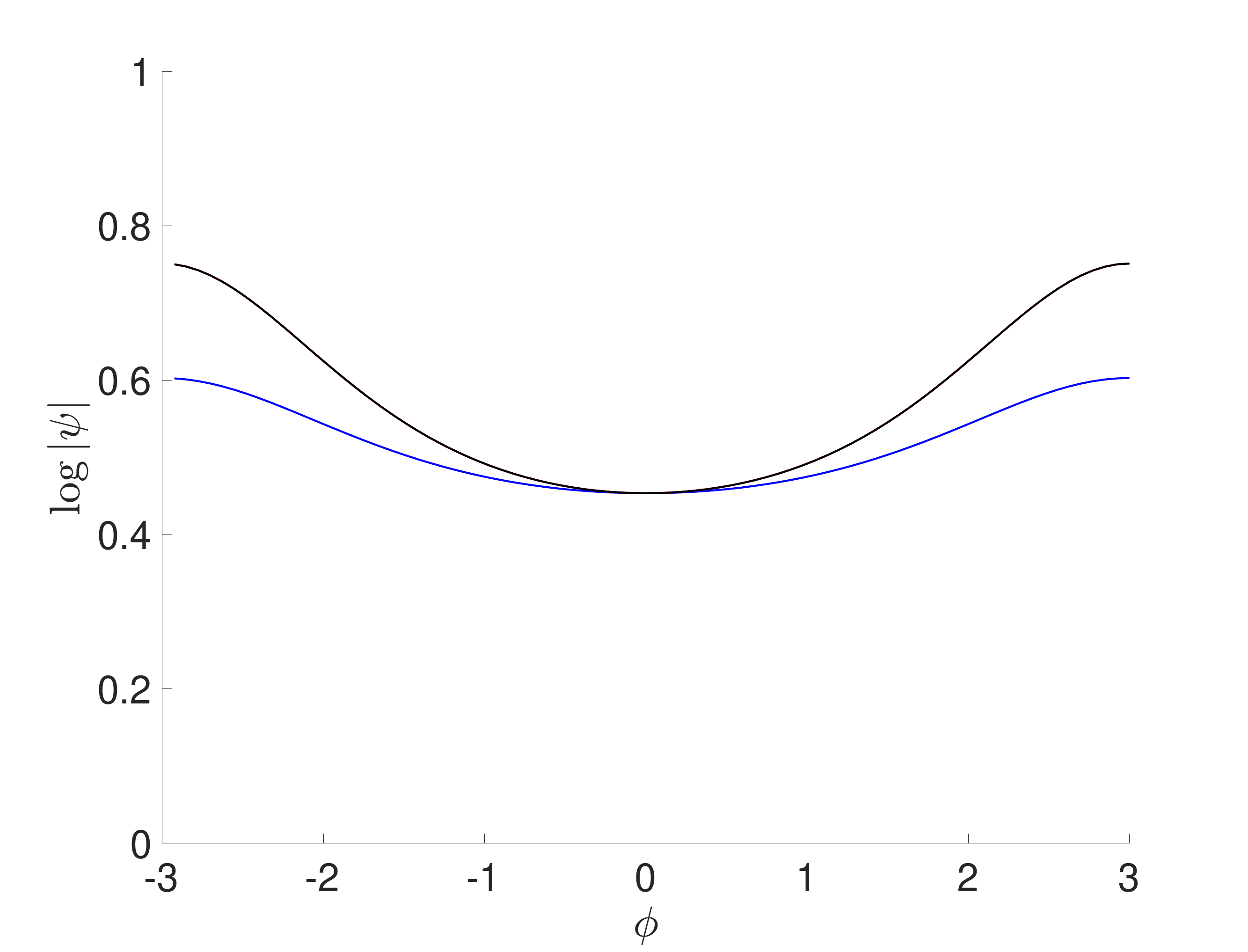}
        \caption{2 dimensional figure.}
        \label{hh_forward_2d_sr}
    \end{subfigure}
    \caption{Evolution of the Wave function $ \psi $ for the no-boundary condition in slow-roll case. The black curve traces the peak of $\psi$ as a function of $\phi$, while the blue curve shows the expected analytical values at the horizon.}
    \label{fig:hh_forward_sr}
\end{figure}

\subsubsection{Tunneling Condition}

In contrast to the no-boundary case, the tunneling proposal requires purely outgoing waves at spatial infinity ($ a \to \infty $) \cite{Vilenkin:1984wp}. To enforce this, we focus $ \psi(x) $ at $ x = 0.95 $ (near the edge of the computational domain) and evolve it backward toward $ x = 0.01 $. This backward propagation suppresses incoming wave components, consistent with the  physical requirements of the tunneling boundary conditions.

\begin{figure}
    \centering
    \begin{subfigure}{0.48\textwidth}
        \includegraphics[width=\textwidth]{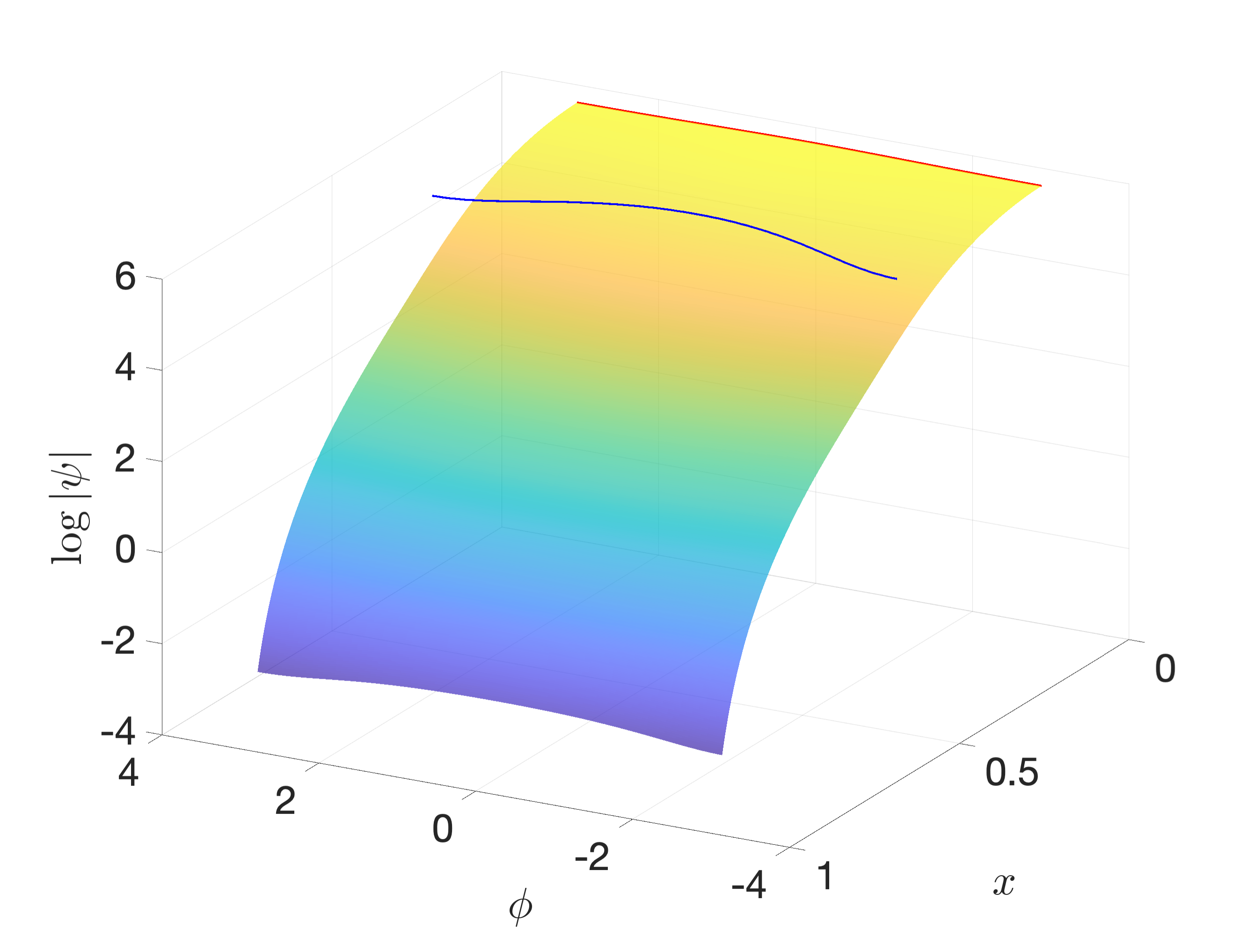}
        \caption{3 dimensional figure.}
        \label{t_backward_sr}
    \end{subfigure}
    \hfill
    \begin{subfigure}{0.48\textwidth}
        \includegraphics[width=\textwidth]{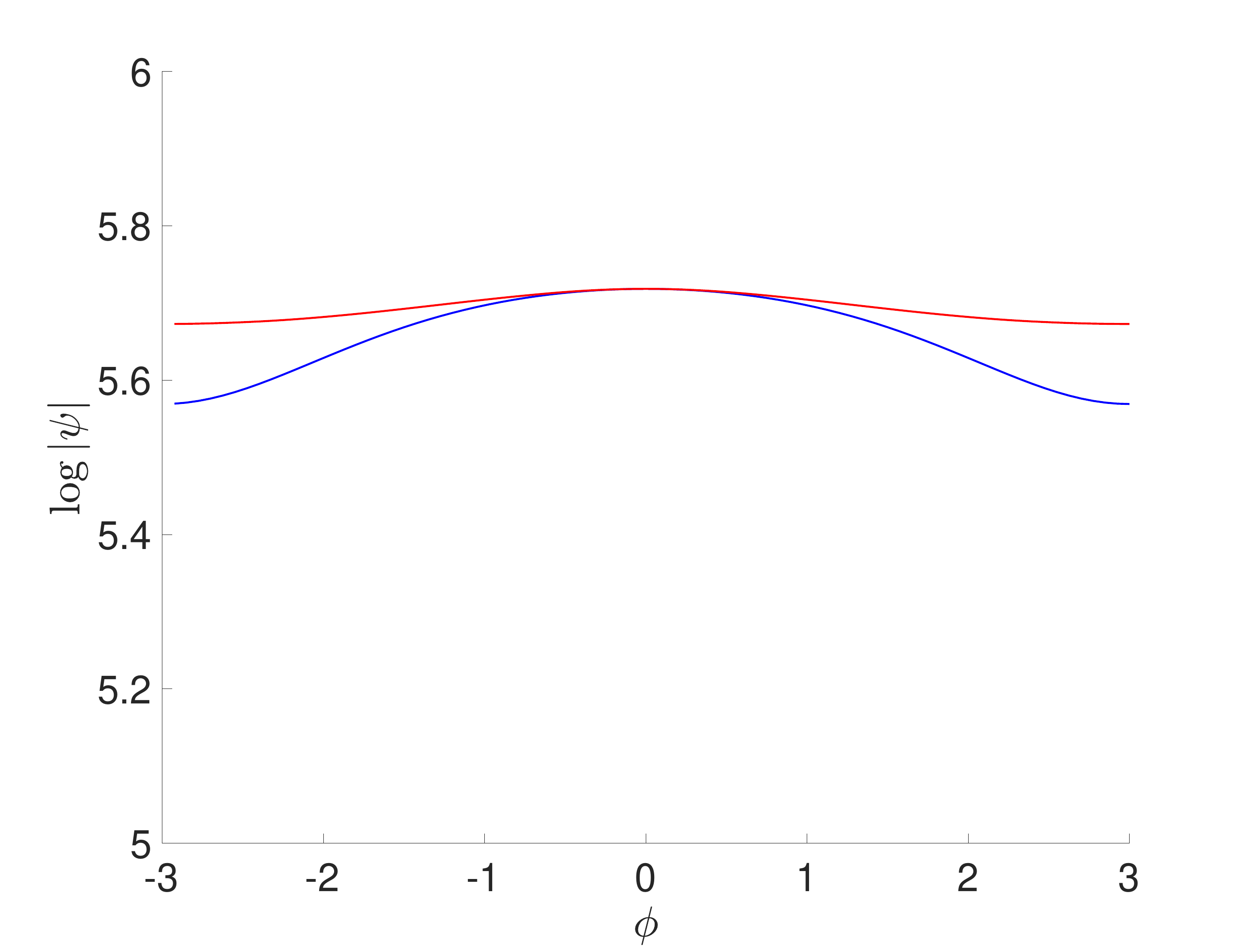}
        \caption{2 dimensional figure.}
        \label{t_backward_2d_sr}
    \end{subfigure}
    \caption{Evolution of the Wave function $ \psi $ for the tunneling condition in slow-roll case. The red curve shows the probability at the initial point $ x _ 0$, while the blue curve indicates the expected analytical solution at horizon.}
    \label{fig:t_backward_sr}
\end{figure}

As depicted in Fig.~\ref{fig:t_backward_sr}, $ \psi(x) $ decreases monotonically as $ x \to 1 $, indicating the exponential decay in a classically forbidden region. The red curve is the probability at the initial point, $ x _ 0 =0.01$. The blue curve is the expected analytical value at horizon. This behavior mirrors quantum tunneling through a potential barrier, where the probability density diminishes exponentially with distance from the turning point. The absence of oscillations reinforces the interpretation of $ x < 0.95 $ as a regime where classical motion is energetically prohibited.  

We note that the  smoothness of our tunneling solution across the entire domain shows the numerical stability of our backward evolution scheme. A comparison with the no-boundary case reveals complementary perspectives.  While the no-boundary condition emphasizes the birth of classical trajectories from a quantum origin, the tunneling condition describes the leakage of probability from a metastable state—a dichotomy central to quantum cosmology.

\subsection{Non-slow-roll Regime ($V_0 = 0.012, m = 0.1, \Delta = 3$)}

\subsubsection{No-boundary Condition}

Figure \ref{fig:hh_forward_nsr} illustrates the two-dimensional wave function $\psi(x, \phi)$ under the no-boundary condition in a non-slow-roll scenario. The horizontal axes are defined by the transformed scale factor $x = \frac{a}{1+a}$ and the field $\phi$, while $\psi$ represents the amplitude of the wave function. The red line indicates the largest values of $\psi$, while the black line marks the second largest values. The blue line represents the expected analytical value at horizon, and it is noteworthy that the black line aligns closely with the horizon line. 

At small values of $x$, the oscillatory behavior of $\psi$ indicates a classical region, while the exponential suppression at large $x$ aligns with semiclassical predictions, illustrating the quantum-to-classical transition of the universe. 

Outside the black line lies the classical region, which exhibits a non-trivial amplification in the non-slow-roll case. Interestingly, this amplification is even more pronounced than the horizon (black line), a distinctive feature of the non-slow-roll scenario.

\begin{figure}
    \centering
    \begin{subfigure}{0.48\textwidth}
        \includegraphics[width=\textwidth]{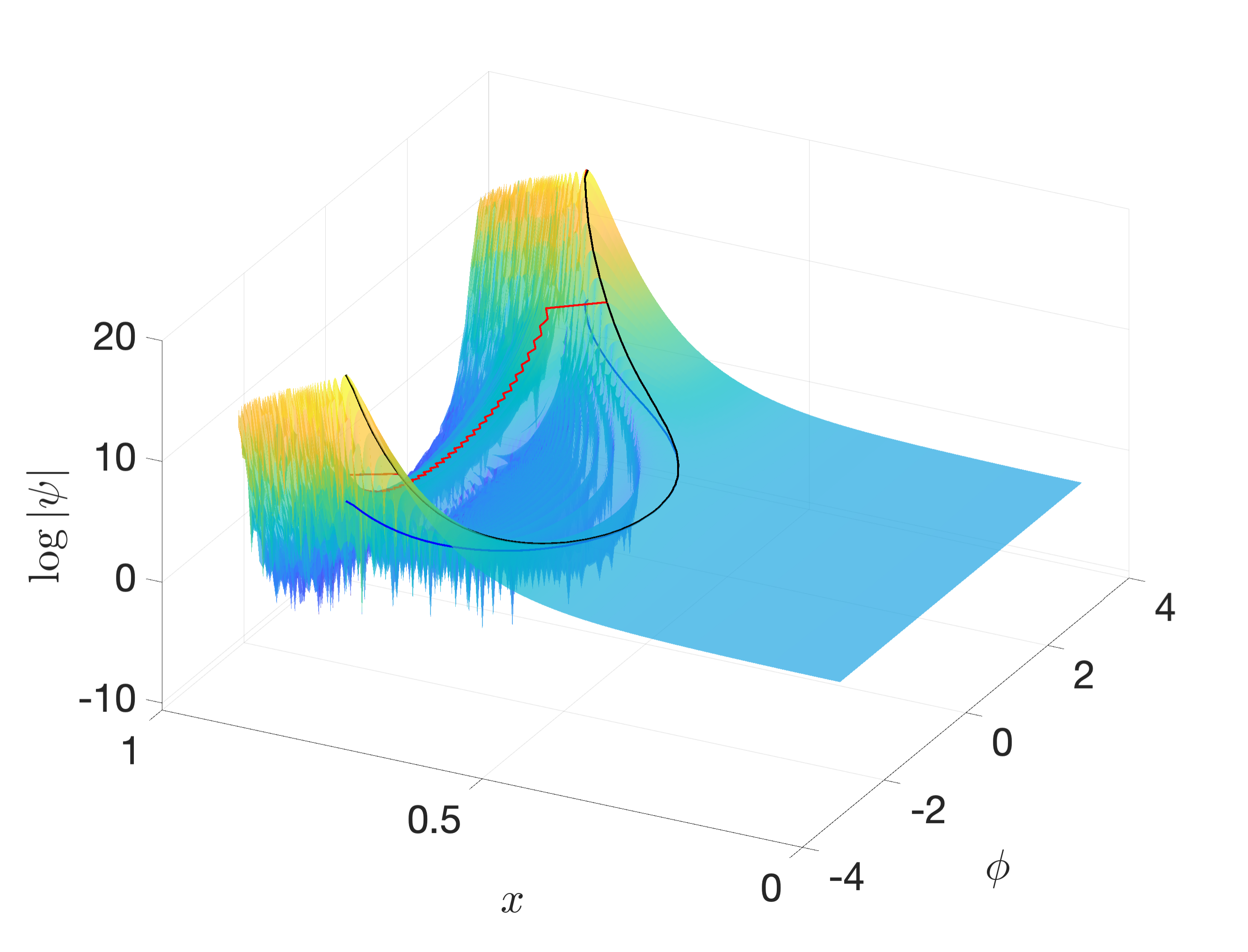}
        \caption{3 dimensional figure.}
        \label{hh_forward_3d_nsr}
    \end{subfigure}
    \hfill
    \begin{subfigure}{0.48\textwidth}
        \includegraphics[width=\textwidth]{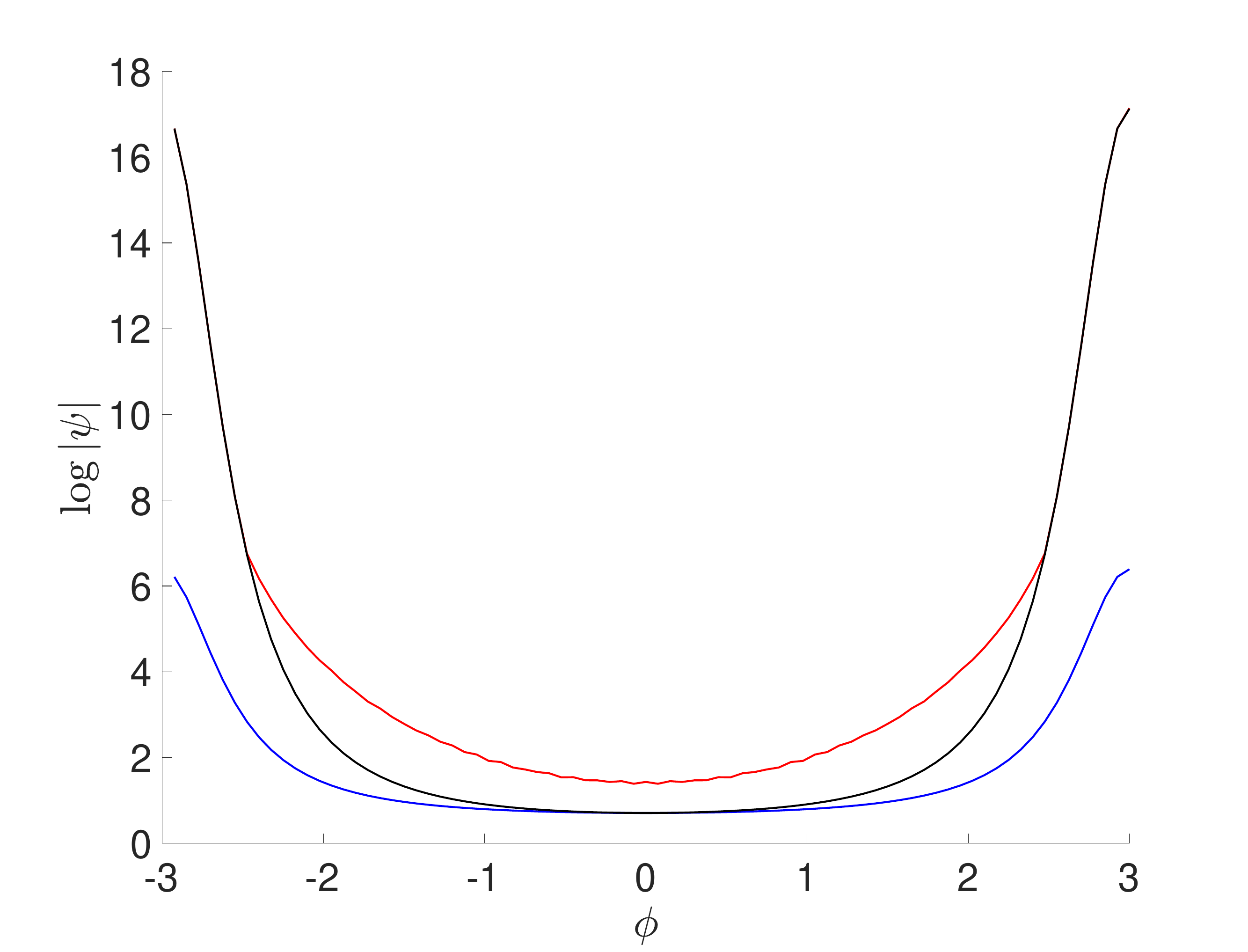}
        \caption{2 dimensional figure.}
        \label{hh_forward_2d_nsr}
    \end{subfigure}
    \caption{Evolution of wave function $ \psi $ for the no-boundary condition in non-slow-roll case. The red line denotes the peak values of $\psi$, the black curve indicated the second maxima, and the blue line shows the expected analytical value at horizon.}
    \label{fig:hh_forward_nsr}
\end{figure}

\subsubsection{Tunneling Condition}

The Fig. \ref{fig:t_backward_nsr} shows the magnitude of the two-dimensional wave function $ |\psi(x, \phi)| $ under the tunneling backward proposal in a non-slow-roll scenario. The red curve marks the wave function very close to the origin, $ x_0 $, the blue curve represents the expected analytical horizon. The behavior of the wave function that decreases as $ x $ decreases while increases as $ \phi $ increases reflects the process of quantum tunneling with non-slow-roll effects causing deviations from classical behavior.

\begin{figure}
    \centering
    \begin{subfigure}{0.48\textwidth}
        \includegraphics[width=\textwidth]{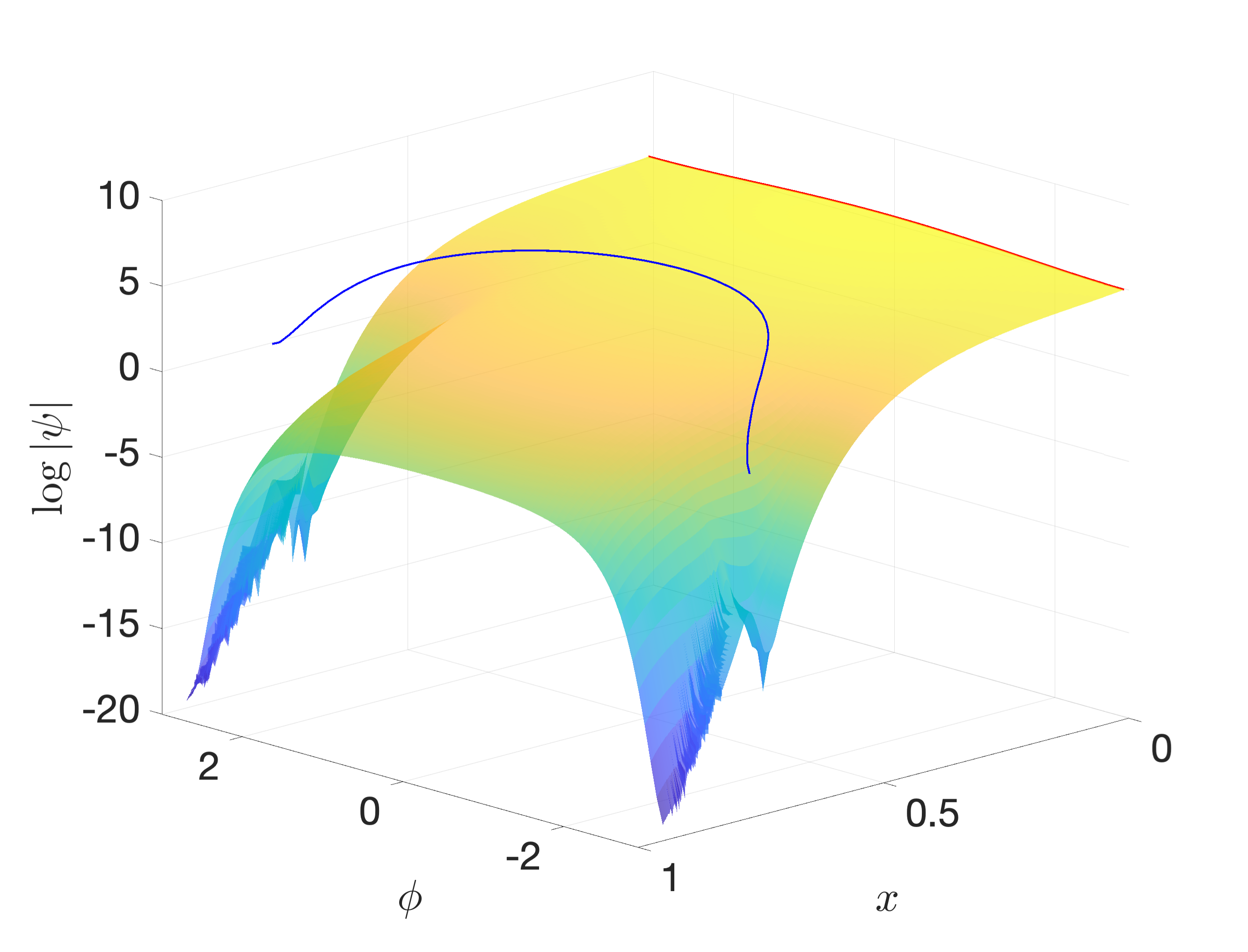}
        \caption{3 dimensional figure.}
        \label{t_backward_3d_nsr}
    \end{subfigure}
    \hfill
    \begin{subfigure}{0.48\textwidth}
        \includegraphics[width=\textwidth]{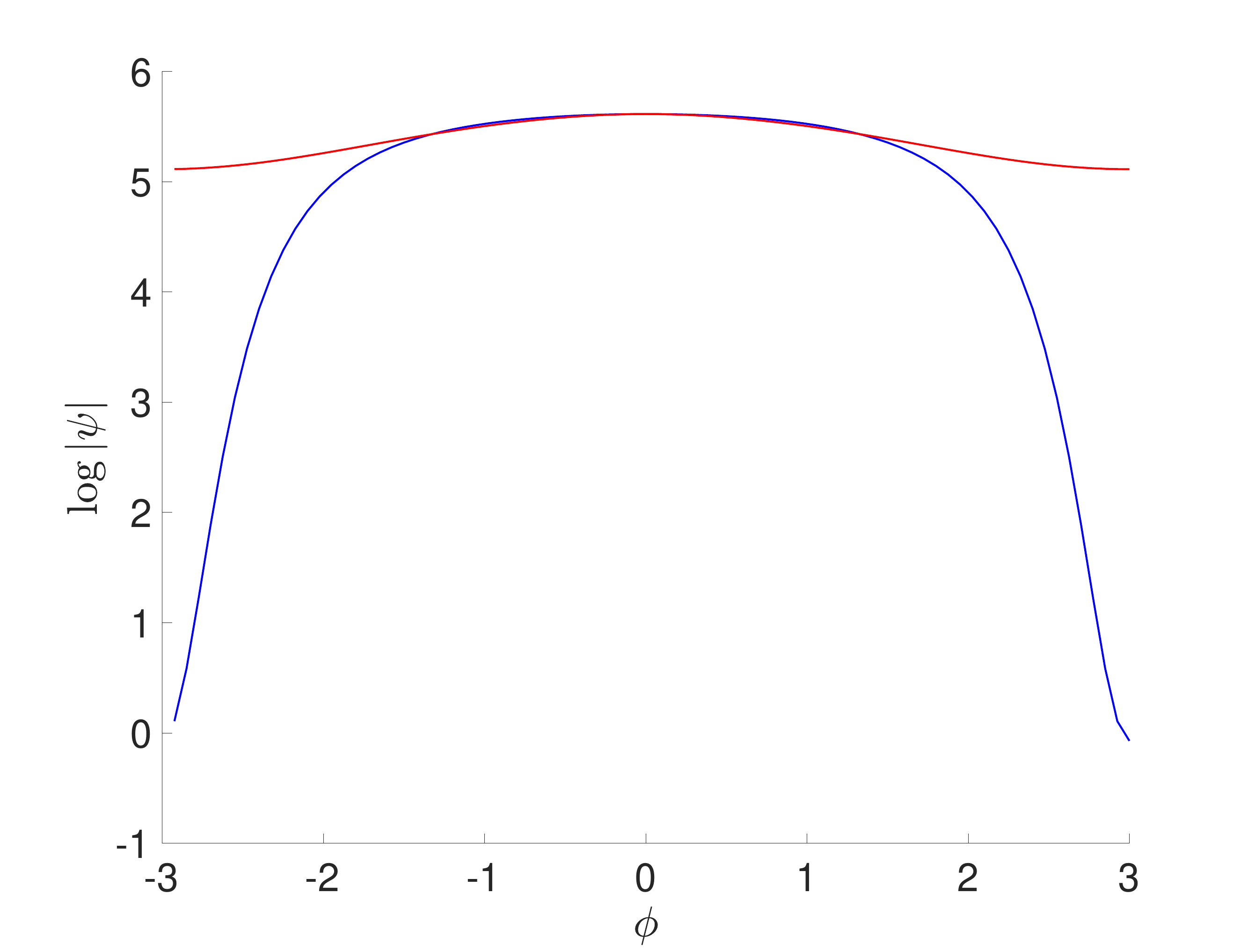}
        \caption{2 dimensional figure.}
        \label{t_backward_2d_nsr}
    \end{subfigure}
    \caption{Evolution of the wave function $ \psi $ for the tunneling condition in non-slow-roll case. The red curve marks $\psi$ near the origin $ x_0 $, while the blue curve corresponds to the expected analytical solution at horizon.}
    \label{fig:t_backward_nsr}
\end{figure}

\subsection{Comparative Analysis}  
The two boundary conditions yield starkly different profiles for $ \psi(x, \phi)$, reflecting their distinct physical interpretations. The no-boundary condition produces a wave function peaked at the horizon, suggesting a ``quantum nucleation" of a classical spacetime. On the other hand, the tunneling condition generates a solution dominated by decay, akin to a metastable vacuum state decaying into a lower-energy configuration.  

These results align with broader debates in quantum gravity, where the choice of boundary conditions often reflects the philosophical stances on the origin of the universe. Future work could quantify tunneling probabilities or explore parameter spaces where both solutions converge, possibly bridging these two frameworks.

\section{Conclusion}

Our analysis elucidates the impact of boundary conditions on the universe’s quantum state. The no-boundary proposal generates a wave function peaked at the horizon, favoring smooth nucleation, while the tunneling condition produces exponentially suppressed solutions to indicate the decay of false vacuum. These differences underscore the challenges in quantum cosmology, where the boundary conditions encode the assumptions about the nature of time and the existence of our universe. 

Our key findings include:
Firstly we find a coordinate transformation from the scale factor to a compact  variable, $x$, which  successfully regularizes the WDW equation, enabling robust numerical solutions.
Secondly, we have analyzed the dependence of the wave function on the potential : Slow-roll parameters modulate the quantum-to-classical transition, while non-slow-roll scenarios exhibit somewhat prolonged quantum effects.
Lastly, we suggest a few observational signatures that the oscillatory behavior in the classical regime may encode fingerprints of the primordial potential, testable through cosmological correlations.

Future work should explore higher-dimensional minisuperspaces, incorporate inhomogeneities, and quantify the tunneling probabilities. Bridging the gap between these proposals remains a critical challenge, potentially achievable through a unified framework reconciling their philosophical underpinnings.

\acknowledgments

This work was supported by the National Research Foundation of Korea (NRF) grant funded by the Korea government (MSIT) (2021R1A4A5031460). 
The authors would like to thank Wan-Il Park for the useful discussion. DY was further supported by the National Research Foundation of Korea (Grant No. : 2021R1C1C1008622) and DKH acknowledges support from Basic Science Research Program through
the National Research Foundation of Korea (NRF) funded by the Ministry of Education (NRF-2017R1D1A1B06033701).



\begin{thebibliography}{200}

\bibitem{DeWitt:1967yk}
B.~S.~DeWitt,
Phys. Rev. \textbf{160}, 1113-1148 (1967).

\bibitem{Hartle:1983ai}
J.~B.~Hartle and S.~W.~Hawking,
Phys. Rev. D \textbf{28}, 2960-2975 (1983).

\bibitem{Vilenkin:1984wp}
A.~Vilenkin,
Phys. Rev. D \textbf{30}, 509-511 (1984).

\bibitem{Hartle:2008ng}
J.~B.~Hartle, S.~W.~Hawking and T.~Hertog,
Phys. Rev. D \textbf{77}, 123537 (2008).
[arXiv:0803.1663 [hep-th]].

\bibitem{Hwang:2013nja}
D.~Hwang and D.~Yeom,
JCAP \textbf{06}, 007 (2014).
[arXiv:1311.6872 [gr-qc]].

\bibitem{Hwang:2012bd}
D.~Hwang, S.~A.~Kim, B.~H.~Lee, H.~Sahlmann and D.~Yeom,
Class. Quant. Grav. \textbf{30}, 165016 (2013).
[arXiv:1207.0359 [gr-qc]].

\bibitem{Maldacena:2024uhs}
J.~Maldacena,
[arXiv:2403.10510 [hep-th]].
\bibitem{Feldbrugge:2017kzv}
J.~Feldbrugge, J.~L.~Lehners and N.~Turok,
Phys. Rev. D \textbf{95}, no.10, 103508 (2017).
[arXiv:1703.02076 [hep-th]].

\bibitem{Brahma:2020cpy}
S.~Brahma, R.~Brandenberger and D.~Yeom,
JCAP \textbf{10}, 037 (2020).
[arXiv:2002.02941 [hep-th]].

\bibitem{Vilenkin:1987kf}
A.~Vilenkin,
Phys. Rev. D \textbf{37}, 888 (1988).

\bibitem{Hwang:2011mp}
D.~Hwang, H.~Sahlmann and D.~Yeom,
Class. Quant. Grav. \textbf{29}, 095005 (2012).
[arXiv:1107.4653 [gr-qc]].

\bibitem{Yeom:2021twr}
D.~Yeom,
Universe \textbf{7}, no.10, 367 (2021).
[arXiv:2106.10790 [gr-qc]].

\bibitem{Kang:2022tkb}
S.~Kang, W.~Park and D.~Yeom,
Symmetry \textbf{16}, no.4, 444 (2024).
[arXiv:2208.12380 [gr-qc]].

\bibitem{Kiefer:2007ria}
C.~Kiefer, ``\textit{Quantum gravity},'' Oxford University Press (2004).

\end{thebibliography}


\end{document}